\documentclass[pra,twocolumn,10pt, superscriptaddress,notitlepage,showpacs]{revtex4-1}
\usepackage{amsmath,amsfonts,amssymb,amstext,amscd,amsthm,dsfont,bbm,natbib}
\usepackage{physics}
\usepackage{color}
\usepackage{soul,xcolor}
\usepackage{gensymb}
\usepackage[colorlinks, breaklinks, urlcolor={cyan}, linkcolor={red}, citecolor={blue}]{hyperref}
\newtheorem{theorem}{Theorem}
\newtheorem{proposition}{Proposition}
\usepackage[normalem]{ulem}
\usepackage{graphicx}
\usepackage{bbold}
\usepackage{braket}
\usepackage{placeins,comment}
\usepackage{caption}
\usepackage{subcaption}
\usepackage{ragged2e}

\newcommand{\stretchket}[1]{\left| #1 \right\rangle}

\renewcommand{\min}{{\mathrm min}}
\renewcommand{\ketbra}[3]{|#1\rangle_{#2}\langle#3|}

\expandafter\let\csname equation*\endcsname\relax
\expandafter\let\csname endequation*\endcsname\relax

\usepackage{graphicx}
\usepackage{hyperref}
\usepackage{float}
\usepackage[normalem]{ulem}
\usepackage{epstopdf}
\usepackage{color}
\begin{document} 

\title{Decoherence-free subspaces and Markovian revival of genuine multipartite entanglement in a dissipative system}
\author{Shubhodeep Gangopadhyay}
\affiliation{Theoretical Sciences Division,
Poornaprajna Institute of Scientific Research (PPISR), 
Bidalur post, Devanahalli, Bengaluru 562164, India}
\affiliation{Graduate Studies, Manipal Academy of Higher Education, Madhava Nagar, Manipal 576104, India}
\author{Vinayak Jagadish}
\affiliation{Department of Computer Science and Engineering, Amrita School of Computing, Amrita Vishwa Vidyapeetham, Amritapuri 690525, India}
\affiliation{National Institute for Theoretical and Computational Sciences (NITheCS), Stellenbosch, 7600, South Africa}
\affiliation{Centre for Quantum Science and Technology, Chennai Institute of Technology, Chennai, 600069, India} 
				\author{R. Srikanth}
\affiliation{Theoretical Sciences Division,
Poornaprajna Institute of Scientific Research (PPISR), 
Bidalur post, Devanahalli, Bengaluru 562164, India}


\begin{abstract} 
We investigate the dynamics of genuine multipartite entanglement (GME) in a system of $n$ qubits ($n\ge3$) collectively interacting with a common zero-temperature bosonic bath characterized by a Lorentzian spectral density. Restricting the dynamics to the single-excitation sector, the collective system–bath coupling naturally separates the Hilbert space into a superradiant mode and a subspace of states orthogonal to it, which forms a decoherence-free (subradiant) subspace. We show that this symmetry-induced structure leads to persistent components of the state that remain protected from dissipation. Specifically, in the three-qubit case, the time evolution of genuine tripartite entanglement is analyzed using the convex roof extension of negativity. We identify parameter regimes determined by the bath spectral density and collective coupling strengths that correspond to Markovian and non-Markovian dynamics. In the Markovian limit, we demonstrate that GME can exhibit a nontrivial revival even in the absence of environmental memory effects. This revival arises from the destructive interference between the decaying superradiant component and the invariant subradiant subspace under suitable system configuration, leading to a transient loss of GME.
\end{abstract}

\maketitle  
\color{black}
\section{Introduction}
In quantum mechanics, the interaction between a system and its environment leads to decoherence, whereby pure quantum states are transformed into statistical mixtures, effectively affecting quantum superpositions and leading to decay of entanglement~\cite{Paz2008,Contreras-Pulido2008, Wahyu_Utami2008,Ma2012, Su2014, Badveli2020}. This phenomenon is described by the theory of open quantum systems~\cite{Davies1976-ab,Haroche2006-jo,Jagadish2018-ya}. The study of open quantum systems has become a cornerstone of modern quantum physics, with profound implications and applications in quantum information science, quantum optics, quantum thermodynamics, including many others. Unlike closed systems that evolve unitarily according to the Schr{\"o}dinger equation, open systems are subject to decoherence and dissipation due to their coupling with an external environment. These effects are crucial in determining the feasibility of quantum technologies such as quantum computing, quantum communication, quantum sensing, etc., where maintaining coherence and controlling entanglement dynamics are essential~\cite{Chuang1995,Shor1995,Aschauer2002,Matsuzaki2011,Albash2015,Schlosshauer-Selbach2007}. 

Systems consisting of multiple qubits interacting with a common bosonic environment have attracted significant attention due to their relevance in cavity quantum electrodynamics (cavity QED)~\cite{Hagley1997,Rauschenbeutel2000,Zheng2000,Rosseau2014,Yang2004,Rogers2017}, circuit QED~\cite{Blais2021} etc. When multiple qubits are coupled to the same electromagnetic field mode, collective effects such as superradiance and subradiance emerge, leading to rich and complex dynamics. Moreover, the nature of the environment, whether it induces Markovian (memoryless) or non-Markovian (memory-preserving) dynamics, plays a critical role in the evolution of the system. Understanding these dynamics is essential for harnessing quantum coherence and entanglement in practical applications.
Recent works have explored new methods of characterizing and protecting coherence in multi-qubit systems, highlighting the critical role of environmental interactions in the preservation of entanglement and the scaling of quantum information processing tasks~\cite{Chiu2025,Gautam2025}. 
A passive strategy to combat decoherence in multi-qubit systems is the use of decoherence-free subspaces (DFS)~\cite{Lidar1998}. These are special subspaces of the Hilbert space that remain invariant under the system-environment interaction, thereby protecting quantum information from external noise. DFS arises due to the symmetry in the coupling between the qubits and the environment. For instance, if multiple qubits interact identically with a common environment, certain entangled states may experience no decoherence, as the collective decoherence mechanisms cancel out.

Recent developments have also focused on the role of non-Markovian dynamics in the evolution of open quantum systems. It is now widely acknowledged that environments with memory can lead to fascinating phenomena, such as entanglement revival, which is not possible in Markovian systems. These non-Markovian effects, where information flows back from the environment to the system, provide a pathway to overcome some of the limitations of traditional quantum error correction techniques. For example, non-Markovian reservoirs have been shown to enhance the fidelity of quantum gates and can be harnessed to improve entanglement-based quantum communication protocols~\cite{Filenga2020} and design more efficient quantum thermal machines~\cite{Lucio2025}. In a similar vein, in non-Markovian environments \cite{breuer_colloquium:_2016, de_vega_dynamics_2017, li_concepts_2017, rivasreview}, entanglement may revive after a period of decay due to backflow of information from the bath to the system.

Motivated by these physical platforms, it is natural to model the environment as a structured bosonic reservoir that captures the spectral properties of cavity or circuit modes. In particular,
a Lorentzian spectral density provides an effective description of a leaky cavity interacting with atoms or artificial qubits, where the spectral width is determined by the cavity damping rate.
This model has the advantage that it allows for an exact treatment of the system dynamics while retaining the essential physics of collective dissipation and reservoir memory effects. In the following, we therefore consider a system of three or more qubits collectively coupled to a bosonic bath with Lorentzian spectral density and analyze the resulting entanglement dynamics\color{black}. This form of spectral density is experimentally realizable in cavity QED setups and photonic bandgap materials, making it a common choice for theoretical studies \cite{Lambropoulos2000, Lewenstein1988,Xu2019}. The interplay between Markovian and non-Markovian effects, especially in the context of multi-qubit systems and collective coupling, is essential for the development of scalable and resilient quantum technologies, which is the focus of the present work. To quantify tripartite entanglement, we use negativity (derived from the partial transpose of the density matrix) in this work.  For a three-qubit system under a common bath, the dynamics of negativity  can reveal transitions between Markovian and non-Markovian regimes, and of special interest, transitions between multi-qubit and fewer-qubit entanglement leading to entanglement revival even in the Markovian regime. The distinction between Markovian and non-Markovian dynamics is not merely academic; it has practical implications for quantum control strategies and information processing~\cite{Bernardes2015-fm, Goswami2021-tn,Rojas-Rojas2024-ml,gulati2024}.

The model of multiple qubits interacting with a common Lorentzian bath is directly applicable to cavity QED, where atoms (or artificial atoms such as superconducting qubits) interact with a single cavity mode. The collective coupling leads to phenomena like superradiance, where the emission rate scales with the square of the number of qubits, and subradiance, where certain states become dark and do not decay. These effects are fundamental not only to quantum optics but also have applications in quantum networking and light-matter interfaces~\cite{Zhong2021}. In quantum thermodynamics, such systems are studied to understand heat transport, work extraction, and the role of coherence in thermodynamic processes~\cite{Hewgill2018,Grimaudo2024}. The interplay between entanglement generation and bath-induced dissipation is crucial for designing quantum thermal machines that outperform their classical counterparts.

It is useful to clarify how the present work relates to earlier studies on environment-induced entanglement revival. In particular, the collective dissipative system studied here generalizes to three and higher number of qubits as opposed to the corresponding two-qubit system studied in~\cite{Maniscalco_2008}. In such systems, the underlying dynamics originates from the coexistence of superradiant and subradiant components in the single-excitation sector governed by a collective system–bath dissipative interaction. In Ref. \cite{Maniscalco_2008}, the focus is on two-qubit concurrence revival due to non-Markovian memory effects, which can be understood in terms of information backflow. By contrast, here our focus is on the revival of genuine-multipartite entanglement observed in the Markovian regime, which is not related to information backflow but arises from the competition between superradiant and subradiant probability amplitudes during the decay. Furthermore, relative to earlier analyses such as Ref.~\cite{Badveli2020}, the present work emphasizes the role of the DFS structure in governing the dynamics of genuine multipartite entanglement, quantified
here through the convex-roof extension of negativity. This perspective allows us to highlight the interplay between collective dissipation, DFS structure, and multipartite entanglement dynamics in both Markovian and non-Markovian regimes. \color{black}

In Sec.~{\ref{model}}, we present the theoretical model and derive the exact dynamics of the system. We then discuss the emergence of decoherence-free subspaces. This is followed by the study of entanglement dynamics using negativity where we explore the transition between Markovian and non-Markovian regimes  in Sec.~{\ref{entanglement}}. The details of the structure of DFS in the $n$-qubit case is addressed in Sec.~{\ref{sec:dfsnqubit}}. This is followed by an important discussion on the revival of entanglement in Sec.~{\ref{sec:revival}}. Finally, we summarize our results and discuss their implications. 

\section{Model and Dynamics}
\label{model}

\subsection{System and equations of motion}
We consider a system of three qubits interacting with a common zero-temperature
bosonic bath. The total Hamiltonian is
\begin{equation}
    H = H_0 + H_{\mathrm{int}},
\end{equation}
where $H_0$ describes the free Hamiltonians of the qubits and the reservoir, and
$H_{\mathrm{int}}$ describes the qubit--bath interaction. Explicitly,
\begin{equation}
\label{hamiltonianfree}
    H_0 = \omega_0 \sum_{i=1}^3 \sigma^{(i)}_{+} \sigma^{(i)}_{-}
    + \sum_k \omega_k b_k^\dagger b_k ,
\end{equation}
\begin{equation}
\label{hamiltonianinteract}
    H_{\mathrm{int}}
    =
    \Big( \alpha_1 \sigma^{(1)}_{+}
        + \alpha_2 \sigma^{(2)}_{+}
        + \alpha_3 \sigma^{(3)}_{+} \Big)
    \sum_k g_k b_k
    + \text{h.c.}
\end{equation}
Here $b_k$ and $b_k^{\dagger}$ are the annihilation and creation operators
of bath mode $k$, $\sigma^{(i)}_{\pm}$ are the raising and lowering operators of
qubit $i$, $\omega_0$ is the common qubit transition frequency, and
$\alpha_i$ denotes the coupling constant of qubit $i$ to the reservoir.

The qubit-bath interaction is characterized by a Lorentzian spectral density
\begin{equation}
\label{specdens}
    J(\omega)
    =
    \frac{1}{\pi}
    \frac{\lambda \gamma}
         {(\omega - \omega_0)^2 + \gamma^2},
\end{equation}
where $\lambda$ is proportional to the vacuum Rabi frequency and $\gamma$
is the spectral width.

We assume the system is initially in the state
\begin{equation}
\label{initialstate}
    \ket{\psi(0)}
    =
    \ket{\psi_S(0)} \otimes \ket{\mathbf{0}}_R ,
\end{equation}
where the qubit state contains a single excitation,
\begin{equation}
\label{sysinit}
    \ket{\psi_S(0)}
    =
    c_1(0)\ket{egg}
    + c_2(0)\ket{geg}
    + c_3(0)\ket{gge},
\end{equation}
and $\ket{\mathbf{0}}_R$ denotes the vacuum of the reservoir.
At time $t$, the general state in the one-excitation sector is
\begin{align}
\label{sbfinal}
    \ket{\psi(t)}
    =&\;
    c_1(t)\ket{egg}\ket{\mathbf 0}_R
    + c_2(t)\ket{geg}\ket{\mathbf 0}_R
    + c_3(t)\ket{gge}\ket{\mathbf 0}_R \notag \\
    &\;
    + \sum_k c_k(t)\ket{ggg}\ket{\mathbf{1}_k}_R ,
\end{align}
where $\ket{\mathbf{1}_k}_R$ denotes a single excitation in mode $k$.

Under the rotating-wave approximation (RWA), the Schr{\"o}dinger equation yields the
equations of motion
\begin{eqnarray}
\label{eom}
    \dot{c}_j(t)
    &=&
    -i \alpha_j \sum_k g_k \, e^{i(\omega_0 - \omega_k)t} \, c_k(t),
    \qquad j=1,2,3,
    \nonumber \\
    \dot{c}_k(t)
    &=&
    -i g_k^* \sum_{j=1}^{3}
    \alpha_j \, e^{-i(\omega_0 - \omega_k)t} \, c_j(t).
\end{eqnarray}

Integrating the equation for $c_k(t)$ and substituting back into the equations
for $c_j(t)$ yields the memory-kernel form
\begin{equation}
\label{eq:cj}
    \dot{c}_j(t)
    =
    - \alpha_j \sum_{l=1}^3 \alpha_l
    \int_0^t dt'\;
    f(t - t') \, c_l(t'),
\end{equation}
where the reservoir correlation function is
\begin{equation}
\label{corrfun}
    f(t - t')
    =
    \sum_k |g_k|^2
    e^{i(\omega_0 - \omega_k)(t - t')}.
\end{equation}

In the continuum limit, using the spectral density~\eqref{specdens},
\begin{equation}
\label{corrfunspec}
    f(t - t')
    =
    \int_{-\infty}^{\infty} d\omega \,
    J(\omega) \,
    e^{i(\omega_0 - \omega)(t - t')}.
\end{equation}
For the Lorentzian $J(\omega)$, this evaluates to
\begin{equation}
\label{corrfuncfinal}
    f(t-t') = \lambda e^{-\gamma |t - t'|}.
\end{equation}
Thus the equations of motion become
\begin{equation}
\label{eomcjfinal}
    \dot{c}_j(t)
    =
    - \alpha_j \sum_{l=1}^3 \alpha_l
    \int_0^t dt'\,
    \lambda e^{-\gamma(t - t')} c_l(t').
\end{equation}
\color{black}
It is convenient to define the collective (superradiant) amplitude
\begin{equation}
\label{superamp}
    c_+(t)
    =
    \sum_{j=1}^3 r_j c_j(t),
    \qquad
    r_j = \frac{\alpha_j}{\alpha_T},
    \qquad
    \alpha_T = \sqrt{\sum_{j=1}^3 \alpha_j^2}.
\end{equation}
Using Eq.~\eqref{eomcjfinal}, one finds the closed equation
\begin{equation}
\label{eomsuperamp}
    \dot{c}_+(t)
    =
    - \lambda \alpha_T^2
      \int_0^t dt'\,
      e^{-\gamma (t - t')} c_+(t') .
\end{equation}
Solving this integro-differential equation yields
\begin{equation}
\label{solsuperamp}
    c_+(t)
    =
    c_+(0) e^{-\gamma t/2}
    \left[
    \cosh\!\left( \frac{\Omega t}{2} \right)
    +
    \frac{\gamma}{\Omega}
    \sinh\!\left( \frac{\Omega t}{2} \right)
    \right],
\end{equation}
where $\Omega = \sqrt{\gamma^2 - 4\lambda \alpha_T^2 }.$
In the limit $\gamma \to 0$, corresponding to an infinite-memory reservoir (``good
cavity''), the correlation function becomes constant,
$f(t - t') = \lambda$.  
In the opposite limit $\gamma \to \infty$ (``bad cavity''), one obtains
$f(t - t') \to (\lambda/\gamma)\delta(t - t')$, and therefore
\begin{equation}
    c_+(t) = c_+(0)\, e^{-\Gamma t}, \qquad
    \Gamma = \frac{\lambda \alpha_T^2}{\gamma},
\end{equation}
which is the Markovian exponential decay rate.
\subsection{Decoherence Free Subspaces}
Decoherence-free subspaces (DFS) provide a passive mechanism for protecting quantum information from environmental decoherence. The basic idea is that certain subspaces of the system Hilbert space remain invariant under the system–environment interaction, so that
states encoded in these subspaces do not become entangled with the
reservoir during the evolution. Formally, a DFS is a subspace $\mathcal{H}_{\mathrm{DFS}}\subseteq
\mathcal{H}_S$ such that the system–reservoir interaction acts
trivially on it. In particular, for every state
$|\psi\rangle\in\mathcal{H}_{\mathrm{DFS}}$ we require that
\begin{equation}\label{eq:DFS_def}
H_{SR}\big(|\psi\rangle\otimes|\xi\rangle_R\big)
=
|\psi\rangle\otimes|\xi'\rangle_R ,\end{equation} where $|\xi\rangle_R$ and $|\xi'\rangle_R$ are states of the reservoir. This condition implies that the system state $|\psi\rangle$ is left
unchanged by the interaction, and no system–environment entanglement
is generated. Consequently, the reduced system state evolves unitarily within $\mathcal{H}_{\mathrm{DFS}}$, making the subspace immune to the dissipative dynamics induced by the reservoir. In the present model, the collective coupling of the qubits to the common reservoir leads to the emergence of such invariant subspaces. In particular, within the single-excitation sector the dynamics
naturally decomposes into a two-dimensional decoherence-free
subspace spanned by subradiant states, together with an orthogonal
radiatively coupled (superradiant) component.
\begin{equation}
    \ket{\psi_+} \propto r_1\ket{egg}+r_2\ket{geg}+r_3\ket{gge},
    \label{eq:super}
\end{equation}
which is defined as the state maximally coupled to the bath. 
The DFS, in this context also called \textit{subradiant} subspace $\mathcal{H}_{\rm sub}$. Subradiant states are orthogonal to \(|\psi_+\rangle\). One readily determines two subradiant states
\begin{align}
\label{subradstates}
    |\psi_-^1\rangle &= \frac{r_2 |egg\rangle - r_1 |geg\rangle}{\sqrt{r_1^2 + r_2^2}}\\
    |\psi_-^2\rangle &= \frac{r_1 r_3 |egg\rangle + r_2 r_3 |geg\rangle - (r_1^2 + r_2^2) |gge\rangle}{\sqrt{r_1^2 + r_2^2}}
    \label{eq:psi12}
    \end{align}
It is convenient to represent the initial state of the three qubits in terms of these three basis states, $\{\ket{\psi_+},|\psi_-^1\rangle, |\psi_-^2\rangle\} $ as
\begin{equation}
\ket{\psi_S(0)}=\eta_-^1|\psi_-^1\rangle+\eta_-^2|\psi_-^2\rangle+\eta_+\ket{\psi_+}.
\label{eq:subsup}
\end{equation}
The superradiant component decays as \(  \ket{\psi_+}(t) =\Phi(t)\ket{\psi_+(0)}\), as shown in Eq.~(\ref{solsuperamp}) while the subradiant components remain constant and therefore
\begin{equation}
|\psi_S(t)\rangle = \eta_+ \Phi(t)|\psi_+\rangle + \eta_-^1 |\psi_-^1\rangle + \eta_-^2 |\psi_-^2\rangle.
\end{equation}
One can get the probability amplitudes as
\begin{subequations}
\begin{align}
    c_1(t)&=r_1\Phi(t)\eta_++\frac{r_1r_3}{\sqrt{\kappa}}\eta_-^2+\frac{r_2}{\sqrt{\kappa}}\eta_-^1\\
    c_2(t)&=r_2\Phi(t)\eta_++\frac{r_2r_3}{\sqrt{\kappa}}\eta_-^2-\frac{r_1}{\sqrt{\kappa}}\eta_-^1\\
    c_3(t)&=r_3\Phi(t)\eta_+-\sqrt{\kappa}\eta_-^2
\end{align}
\label{eq:3c}
\end{subequations}
where $\kappa=r_1^2+r_2^2$. In the basis $\{\ket{eee},\ket{eeg},\cdots, \ket{ggg}\}$, the reduced three-qubit density matrix $\rho_{123}$ takes the form
\begin{equation}
\label{Eq:RWA}
\rho_{123}= \left(\begin{array}{cccccccc}
0 & 0 & 0 & 0 & 0 & 0 & 0 & 0 \\
0 & 0 & 0 & 0 & 0 & 0 & 0 & 0 \\
0 & 0 & 0 & 0 & 0 & 0 & 0 & 0 \\
0 & 0 & 0 & \left|c_{1}\right|^{2} & 0 & c_{1} c^*_{2} & c_{1} c^*_{3} & 0 \\
0 & 0 & 0 & 0 & 0 & 0 & 0 & 0 \\
0 & 0 & 0 & c_{2} c^*_{1} & 0 & \left|c_{2}\right|^{2} & c_{2} c_{3}^* & 0 \\
0 & 0 & 0 & c_{3} c^*_{1} & 0 & c_{3} c^*_{2} & \left|c_{3}\right|^{2} & 0 \\
0 & 0 & 0 & 0 & 0 & 0 & 0 & \left|c_{k}\right|^{2}
\end{array}\right),
\end{equation}
where the $c_j=c_j(t)$ as obtained above.

\section{Dynamics of Entanglement}
\label{entanglement}
Eq.~(\ref{Eq:RWA}) is now employed for the analysis of the temporal evolution of three-qubit entanglement. Here we will be mainly concerned with genuine multipartite entanglement (GME), rather than the weaker concept of absolute entanglement. GME indicates a quantum state of three or more parties that cannot be described as a mixture of separable states across any bipartite partition. It is a strong form of correlation where the system cannot be produced by combining smaller entangled groups, making all particles essential to the entanglement structure. We refer to a state as \emph{genuinely tripartite entangled} if it cannot be written as a convex mixture of states that are separable with respect to any bipartition of the three subsystems. Among various measures available, a particularly convenient one for our purpose is found to be \textit{tripartite negativity} ($\mathcal{N}_{(3)}(\rho)$), a quantification of pure-state genuine tripartite entanglement. \color{black}This is defined as the geometric mean~\cite{Vidal_2002,Sab_n_2008}
\begin{equation}
   \mathcal{N}_{(3)}(\rho)=(\mathcal{N}_{1:23}\mathcal{N}_{2:31}\mathcal{N}_{3:12})^{\frac{1}{3}},
   \label{eq:tneg}
\end{equation}
where
 \begin{equation}
  \mathcal{N}_{1:23}\equiv\frac{||\rho^{T_1}||-1}{2},   
 \end{equation}
and similarly for $\mathcal{N}_{2:31}$ and $\mathcal{N}_{3:12}$.
Here $||\rho^{T_1}||$ is the trace norm  of the partial transpose of state $\rho_{123}$ with respect to the subsystem $1$ and $\bra{\zeta_1,\zeta_2,\zeta_3}\rho^{T_1}_{123}\ket{\zeta_{\tilde{1}},\zeta_{\tilde{2}},\zeta_{\tilde{3}}}\equiv \bra{\zeta_{\tilde{1}},\zeta_2,\zeta_3}\rho_{123}\ket{\zeta_1,\zeta_{\tilde{2}},\zeta_{\tilde{3}}}$. Since the system evolves into mixed states in general, the measure Eq.~(\ref{eq:tneg}) is not applicable. To cover the general case, we shall use the convex roof extension of negativity (CREN), given by:
\begin{equation}
    \mathcal{N}_{\rm CR}(\rho) =  \min_{\{p_j,\ket{\psi_j}\}} \sum_jp_j\mathcal{N}_{(3)}(\ket{\psi_j}),
\end{equation}
where $\rho = \sum_j p_j\ket{\psi_j}\bra{\psi_j}$. By construction, CREN is convex, reduces to Eq.~(\ref{eq:tneg}) in the pure state case, and is an entanglement monotone, i.e., non-increasing under local operations and classical communication (LOCC). Further, monotonicity of entanglement guarantees that mixing an entangled state with a separable state cannot increase entanglement.

Noting that the total state is given by \begin{equation}
|\psi(t)\rangle= \sqrt{Q(t)}\ket{\widetilde{\psi}_S(t)}_{123}\ket{\bf 0}_R + \sqrt{1-Q(t)}\ket{ggg}_{123}\ket{\bf 1_k}_R,\end{equation}
where $
    Q(t) = |\braket{\psi_S(t)|\psi_S(t)}|^2 = |c_1(t)|^2+ |c_2(t)|^2+ |c_3(t)|^2$ and $\ket{\widetilde{\psi}_S(t)} = \frac{1}{\sqrt{Q(t)}}\ket{\psi_S(t)}$,
a natural decomposition of the reduced system is 
\begin{equation}
    \rho_{123}(t) = Q(t) \ketbra{\widetilde{\psi}_S(t)} {123} {\widetilde{\psi_S}(t)} + [1-Q(t)]\ketbra{ggg}{123}{ggg}.
    \label{eq:canonical}
\end{equation} For these states $\rho_{123}(t)$, evidently an upper bound on CREN can be given:
\begin{equation}
    \mathcal{N}_{\rm CR}(\rho[t]) \le  Q(t) \mathcal{N}_{(3)}\big(\ket{\widetilde{\psi}_S(t)}\bra{\widetilde{\psi}_S(t)}\big) \equiv \mathcal{N}_{\rm CR}^{\ast}(\rho[t]).
    \label{eq:CREN*}
\end{equation} 

Fortunately, it turns out that for the impoverished system we consider (mixture of a $W$-class state and ground state), the decomposition Eq. (\ref{eq:canonical}) of state $\rho_{123}(t)$ is already optimal to CREN. In other words, $\mathcal{N}_{\rm CR}(\rho[t]) = \mathcal{N}_{\rm CR}^{\ast}(\rho[t])$, so that
\begin{equation}
    \mathcal{N}_{\rm CR}(\rho[t]) =  Q(t) \mathcal{N}_{(3)}\big(\ket{\widetilde{\psi}_S(t)}\bra{\widetilde{\psi}_S(t)}\big).
    \label{eq:CREN}
\end{equation}
Before establishing this, we first note using the PPT criterion (which provides a sufficient condition for entanglement) that any state of the form Eq. (\ref{eq:canonical}), i.e., a mixture of a $W$-class state and a vacuum state, is entangled for all $|\eta_{-}^2\big|>0$. Consider the normalized $W$-class state $\ket{\psi_W} \equiv a\ket{egg}+b\ket{geg}+c\ket{gge}$ mixed incoherently with $\ket{ggg}$ in the proportion $p:(1-p)$. Performing partial transpose on the first qubit, we obtain one of four eigenvalues given by $v\equiv1-p-\sqrt{(1-p)^2+4a^2p^2(1-a^2)}$, which is clearly negative for $a,p>0$. A similar result follows partial transposing on the second or third qubit.

However, this only establishes entanglement in the bipartition $A|BC$, etc., but as such does not rule out biseparability. Below we prove GME from algebraic rather than information theoretic arguments, by essentially showing that for our impoverished system (an incoherent mixture of a $W$-class and ground states), biseparability forces mixing, which destroys the rank-1 structure in the single-excitation Hilbert space.
\begin{theorem}
Let $\rho_W = p|\psi_W\rangle\langle\psi_W| + (1-p)|ggg\rangle\langle ggg|$, where
$|\psi_W\rangle = a|egg\rangle + b|geg\rangle + c|gge\rangle$, $a,b,c > 0$,
$a^2+b^2+c^2=1$. Then $\rho$ is genuinely tripartite entangled if and only if $p>0$.
\label{thm:3GME}
\end{theorem}
\begin{proof} The ``if'' direction ($p>0$): Consider $\rho_1 \equiv \rho_W\vert_{\mathcal{H}_1}$, which is $\rho_W$
restricted to the single-excitation
sector $\mathcal{H}_1 \equiv \mathrm{span}\{|egg\rangle,|geg\rangle,|gge\rangle\}$. This yields the pure (but subnormalized) state $\rho_1 = \Pi_{\mathcal{H}_1}\rho_W\Pi_{\mathcal{H}_1} = p\ketbra{\psi_W}{}{\psi_W}$, where $\Pi_{\mathcal{H}_1}$ is the projector to $\mathcal{H}_1$. By virtue of its purity, ${\rm rank}(\rho_1)=1$ and ${\rm det}(\rho_1)=0$. A general pure state $|\phi\rangle = \beta|egg\rangle+\gamma|geg\rangle+\delta|gge\rangle
\in \mathcal{H}_1$ is biseparable across the partition $A|BC$ only if $\beta=0$ or $\gamma=\delta=0$, hence $\beta\gamma = \beta\delta = 0$, i.e., it is in general a mixture of states having the form $\ket{egg}$ or $\mu\ket{geg}+\nu\ket{geg}$. Thus, this component can only contribute coherences of the form $\bra{geg}\rho_W\ket{gge}$. By a similar argument, a biseparable component in the partition $B|AC$ can only yield the coherence $\braket{egg|\rho_W|gge}$, and partition $C|AB$ only $\braket{egg|\rho_W|geg}$. Thus no single biseparable state can produce all the three independent nonvanishing coherences in $\rho_W$, and one must mix states from the three different bipartitions $A|BC, B|AC, C|AB$. For such a mixture $\rho_{\rm bisep}$, we will have rank ${\rm rank}(\rho_{\rm bisep})\ge 2$, and ${\rm det}(\rho_{\rm bisep})>0$, contradicting the corresponding values for $\rho_1$. This rules out any biseparable decomposition of $\rho_W$. 

The ``only if'' direction $(p=0)$: In this case, within this family $\rho_W = \ketbra{ggg}{}{ggg}$, which manifestly lacks (genuine tripartite) entanglement. 
\hfill
\end{proof}
We note that any of the states $\rho_{123}(t)$ in Eq. (\ref{eq:canonical}) correspond to an incoherent mixture of a $W$-class state and $\ket{ggg}$. Then Theorem \ref{thm:3GME} implies that $\rho_{123}(t)$ has GME if and only if $p>0$ in this decomposition, thus establishing the optimality of said decomposition. Therefore it follows that $\mathcal{N}_{\rm CR}$ in this case is indeed given by $\mathcal{N}_{\rm CR}^{\ast}$, i.e., Eq. (\ref{eq:CREN}). \color{black}

To gain deeper insight into how genuine tripartite entanglement, as quantified by $\mathcal{N}_{\rm CR}$, evolves over time based on the system's starting entanglement, we focus on the initial quantum state of the system of the form Eq.~(\ref{sysinit}) with coefficients 
  \begin{align}
  c_1(0)=\sqrt{\dfrac{1+2p}{3}}, c_2(0)=\sqrt{\dfrac{1-p}{3}}e^{i\theta},c_3(0)=\sqrt{\dfrac{1-p}{3}}e^{i\phi}
  \label{eq:coeffinit}
\end{align}
with $0\leq p\leq 1$, where the separability parameter is chosen to represent varying degrees of entanglement ($p=1$ corresponding to a product state, and $p=0$ being the W state, which is the most entangled state in this family). This approach allows us to explore how the initial conditions influence the dynamics of entanglement during the system's evolution.
\begin{figure}[htb!]
    \centering
    \begin{subcaptionbox}{\label{fig:p1}}[0.21\textwidth]
        {\includegraphics[width=\linewidth]{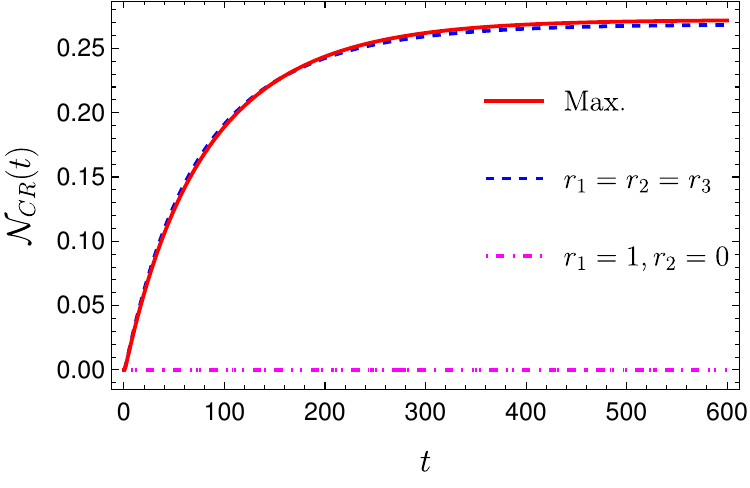}}
    \end{subcaptionbox}
    \hfill
    \begin{subcaptionbox}{\label{fig:p2}}[0.21\textwidth]
        {\includegraphics[width=\linewidth]{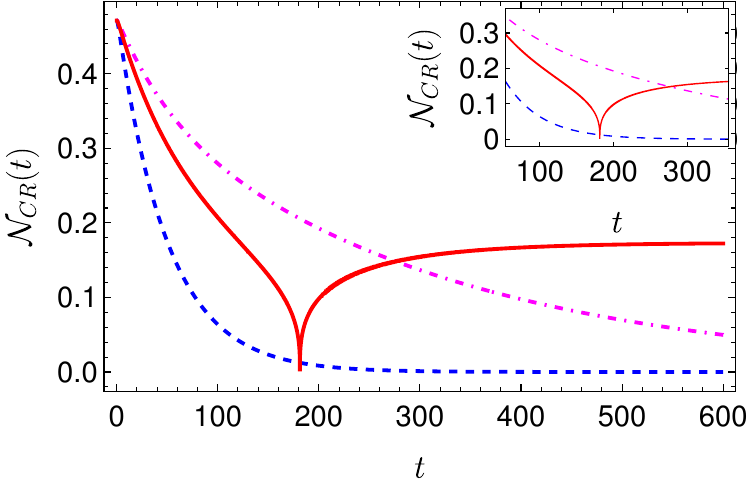}}
    \end{subcaptionbox}

    \medskip

    \begin{subcaptionbox}{\label{fig:p3}}[0.21\textwidth]
        {\includegraphics[width=\linewidth]{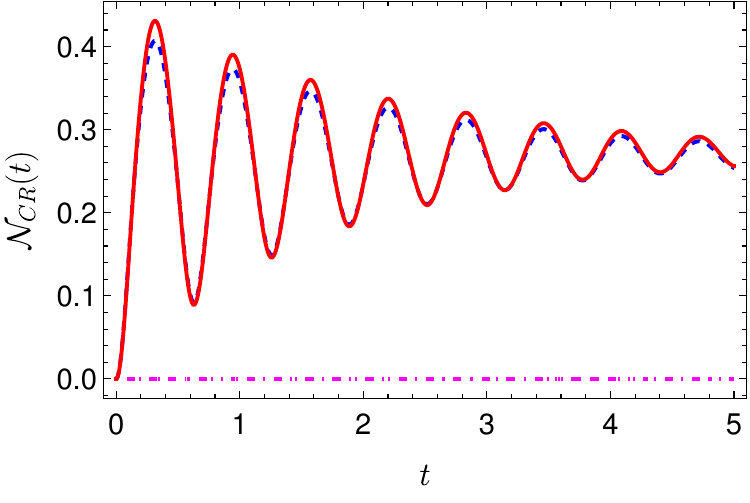}}
    \end{subcaptionbox}
    \hfill
    \begin{subcaptionbox}{\label{fig:p4}}[0.21\textwidth]
        {\includegraphics[width=\linewidth]{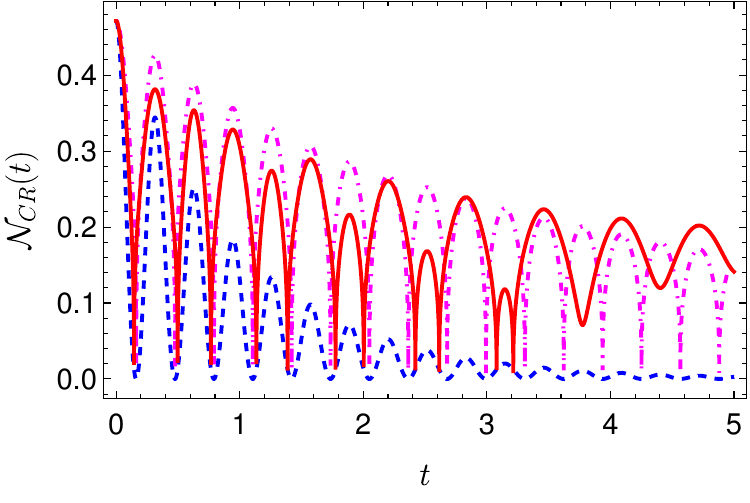}}
    \end{subcaptionbox}

    \caption{(Color Online)\justifying{The dynamics of tripartite negativity are illustrated in both bad cavity limit [i.e. $R=0.1$, top plots (a) and (b)] and good cavity regime [$R=10$, bottom plots (c) and (d)] for two different scenarios (i) with initial separable state, i.e., $p = 1$ in Eq.~(\ref{eq:coeffinit}) [plots (a) and (c)], and (ii) with initial $W$ state $p = 0$ [plots (b) and (d)], keeping both $\theta =0$ and $\phi=0$. Three coupling configurations are examined for all the plots, (i) Symmetric coupling case, $r_1 = r_2 = r_3 = 1/\sqrt{3}$, expressed by dashed lines, (ii) the scenario where only one particle is coupled, $r_1 = 1,r_2 = 0,r_3 = 0$, in dotdashed and (iii) maximum attainable value $r_1=0.53,r_2=0.6$ for $p=1$ and $r_1=0.11,r_2=0.11$ for $p=0$ by solid lines. (The plot labels are the same across all figures, given by the legends in subfigure (a). Here ``Max'' refers to the state that maximizes the asymptotic $\mathcal{N}_{\rm CR}$)} }
    \label{fig:fourplots}
\end{figure}

The evolution of genuine tripartite entanglement measure $\mathcal{N}_{\rm CR}$ as a function of time $t$ is shown in  Fig.~\ref{fig:fourplots}. Fig.~\ref{fig:fourplots}(a) depicts the case of an initially separable state [$p=1$ in Eq.~(\ref{eq:coeffinit})] in the bad cavity ($R =0.1$) limit where $R=r/\gamma$. The solid (red), dashed (blue) and dash-dotted (magenta) plots represent the cases of maximum asymptotic entanglement, uniform coupling ($r_1=r_2=r_3=\frac{1}{\sqrt{3}}$) and $(r_1=1, r_2=r_3=0)$. In the first case (solid curve), the tripartite negativity $\mathcal{N}_{\rm CR}(t)$ starts from zero and increases monotonically owing to the buildup of genuine entanglement induced by the environment until it reaches its asymptotic maximum value ($0.27$). This corresponds to $(r_1,r_2)=(0.53,0.6)$. The stable maximum value reflects the fact that the initial state has finite support in the decoherence-free (subradiant) subspace, Eq.~(\ref{eq:subsup}). Note that the solid and dashed lines are almost coincident owing to the fact that the maximal tripartite negativity is close to the uniform value of ($r_1=r_2=r_3=\frac{1}{\sqrt{3}}$). The dash-dotted plot remains constantly 0, evidently because only one atom is coupled to the environment and thus there is no buildup of tripartite negativity.

Fig.~\ref{fig:fourplots}(b) depicts the same situation as in Fig.~\ref{fig:fourplots}(a), but with an initial $W$ state [$p=0$ in Eq.~(\ref{eq:coeffinit})]. As before, the solid (red), dashed (blue) and dash-dotted (magenta) plots represent the cases of maximum asymptotic entanglement, uniform coupling ($r_1=r_2=r_3=\frac{1}{\sqrt{3}}$) and $(r_1=1, r_2=r_3=0)$. The dashed line corresponds to the case where the initial $W$ state is the superradiant state, and is thus annihilated asymptotically. This also provides the case of the most efficient decoherence, as the dynamical map is essentially an amplitude damping in the qubit subspace given by span[$\ket{\psi^+}, \ket{ggg}$] (refer Section \ref{sec:revival}). The dash-dotted plot represents the case where tripartite entanglement is destroyed through decoherence on one of the particles. The solid curve corresponds to the asymptotic maximum value of $\mathcal{N}_{\rm CR}$. This corresponds to $(r_1,r_2)=(0.11,0.11)$ and yields $\mathcal{N}_{\rm CR}(t=\infty)=0.19$. In this case, the initial tripartite negativity $\mathcal{N}_{\rm CR}(\rho)$ decays to 0 due to the occurrence of a biseparable state, followed by a revival until it reaches its asymptotic value.  In the following, we discuss the origin and genuineness of the death and revival of the tripartite entanglement here. 

The revival in Fig.~\ref{fig:fourplots}(b) would be applicable in this scenario with initial $\ket{W}$ for the case $r_1=r_2$ (more generally, any two coupling constants being equal). In this case note that $\braket{W|\psi_{-}^1} \equiv \eta_{-1}=0$, i.e., the initial state has only a one-dimensional support in the DFS. The initial system-reservoir state is then given by
\begin{equation}
    \ket{\Psi(0)}\ket{\bf 0}_R = \ket{W}\ket{\bf 0}_R = (\eta_+\ket{\psi_+} + \eta_-^2|\psi_-^2\rangle)\ket{\bf 0}_R.
    \label{eq:psi_init}
\end{equation}
 Since the superradiant state is annihilated, and the DFS state remains invariant, the final state will have the form:
\begin{equation}
    \ket{\Psi(\infty)} = \eta_-^2|\psi_-^2\rangle\ket{\bf 0}_R + \eta_+\ket{ggg}\ket{\bf 1_k}_R.
    \label{eq:psi_fin}
\end{equation}
The process of the decoherence essentially transfers the probability with the superradiant state to the $\ket{ggg}$ state. By virtue of continuity, the transformation from Eq.~(\ref{eq:psi_init}) to Eq.~(\ref{eq:psi_fin}) will involve an intermediate state like
\begin{align}
    \ket{\Psi(t)} &= \eta_-^2|\psi_-^2\rangle\ket{\bf 0}_R 
    + \eta_{+}\big(\alpha(t)\ket{\psi_+}\ket{\bf 0}_R \nonumber \\ &+\beta(t)\ket{ggg}\ket{\bf 1_k}_R\big),
\end{align}
where $\alpha(t)$ (resp., $\beta(t)$) is monotonically decreasing to 0 (resp., increasing to 1)  in the bad cavity limit and $|\alpha(t)|^2 + |\beta(t)|^2=1$. The state projected within the single-excitation sector can be expanded as 
\begin{align}
    \ket{\Psi_S(t)^{[1]}} &=\bigg(\dfrac{\eta_-^2}{\sqrt{2}}\sqrt{1-2r_1^2}+\alpha\eta_+r_1\bigg)(\ket{egg} + \ket{geg}) \nonumber \\
&+\bigg(\eta_+\alpha\sqrt{1-2r_1^2}-\sqrt{2}\eta^2_- r_1\bigg)\ket{gge}.
\label{eq:psi_t}
\end{align}
\color{black}
Under continuous evolution, there is a finite time $t^{\star} >0$ such that the amplitude of the state $\ket{gge}$ above momentarily vanishes, i.e., 
$\eta_+\alpha(t^{\star})\sqrt{1-2r_1^2} = \sqrt{2}\eta^2_- r_1$
This is guaranteed provided 
\begin{equation}
    \eta_+\sqrt{1-2r_1^2} > \sqrt{2}\eta_-^2 r_1,\quad \eta_-^2>0.
    \label{eq:tstar}
\end{equation} 
Note that this condition is not satisfied for the case of dashed line, with $r_1=r_2=r_3=\frac{1}{\sqrt{3}}$, since $\eta_-^2=0$ (even though the first condition in Eq.~(\ref{eq:tstar}) is satisfied noted that $\eta_+=1$).

Here it is worth noting how the assumptions $\theta=\phi=0$ are helpful. Rewriting Eq. (\ref{eq:tstar}) explicitly, we have
\begin{equation}
    (2r_1c+r_3c_3)\sqrt{1-2r_1^2} > 2(r_3 c - r_1c_3)r_1,
\label{eq:thetaphi}
\end{equation}
where $c\equiv c_1=c_2$ (as follows from Eq. (\ref{eq:psi12})).
Notice that we can arrange for this to be satisfied by choosing small $c, r$ and large $c_3$ (as is done in our example) when $c,c_3>0$. If $\phi=\pi$, i.e., $c_3 \rightarrow -c_3$, then we find that the conditions to make the LHS large contradict those to make the RHS small. For example, we can consider making $c_3,r_1$ small to make the RHS in this case small. But this also diminishes the LHS, requiring us to make $c$ large, which in turn increases the RHS. This impossibility to satisfy the requirement Eq. (\ref{eq:thetaphi}) can verified numerically, and the corresponding evolution shows no event of entanglement death-revival in its approach to the nonvanishing asymptotic value. \color{black}

Referring to Eq.~(\ref{eq:psi_t}), we find that at time $t^{\star}$ the resultant intermediate state $\ket{\Psi_S(t)^{[1]}} \propto (\ket{eg}+\ket{ge})\ket{g}$, i.e., it becomes bi-separable, so that $\mathcal{N}_{\rm CR}$ vanishes. By virtue of Eq.~(\ref{eq:CREN}), CREN vanishes. As the system evolves beyond $t^{\star}$, once again $\mathcal{N}_{\rm CR}$ rises to the asymptotic value guaranteed by the stationary states, given by the state in Eq.~(\ref{eq:psi_fin}). Moreover, this happens in the Markovian regime, i.e., one where the dynamics is CP-divisible, which is explained later below. Here Eq.~(\ref{eq:canonical}) assumes the asymptotic form:
\begin{equation}
    \rho_{123}(\infty) = \big|\eta_{-}^2\big|^2 \ketbra{\psi_{-}^2} {123} {\psi_{-}^2} + \big|\eta_{+}\big|^2\ketbra{ggg}{123}{ggg}.
    \label{eq:canonical0}
\end{equation}

We note that an analogous entanglement-revival occurs in the corresponding two-qubit case. This effect appears in Figure 2(b) of~\cite{Maniscalco_2008}, but apart from a brief allusion to a similar behavior in Ref. \cite{ficek2006dark}, the authors don't discuss the effect further, since their primary focus is on the largeness of their entanglement revivals in the strong coupling (non-Markovian) regime. Interestingly, in Ref. \cite{ficek2006dark}, the authors report a Markovian revival of entanglement that bears some similarity to our result. They discuss the decay of two two-level atoms an initially double-excited state $\ket{ee}$ through a mixture of the symmetric (superradiant) state $\ket{s} \propto (\ket{ea}+\ket{ae})_{12}$, the antisymmetric (subradiant) state $\ket{a} \propto \ket{ea}-\ket{ae}$ state and the ground state $\ket{gg}$, in addition to $\ket{e}  \equiv \ket{e}_1\ket{e}_2$. The evolution experiences a sequence of two entanglement deaths and rebirths before eventually decaying to the ground state $\ket{g} \equiv \ket{g}_1\ket{g}_2$. It turns out that the second rebirth there is somewhat similar to the entanglement revival in our case. However, there are important differences. 

Foremost, our result applies to GME in multi-qubit systems, whereas there it is restricted to bipartite entanglement. In our case as well as that of Ref. \cite{Maniscalco_2008}, the ``dark phase'' (where genuine tripartite entanglement vanishes) is momentary, whereas in Ref. \cite{ficek2006dark} the dark phase (where bipartite entanglement vanishes) persists for finite time. This is essentially because the interim mixed state in our case contains only the ground state and a single-excitation state, whereas there the state $\ket{ee}$ is an extra. Around the second rebirth event reported in Ref. \cite{ficek2006dark}, the concurrence $\mathcal{C} \equiv {\rm max}\{0, |P_{ss}-P_{aa}|-2\sqrt{P_{ee}P_{gg}}\}$ for this class of states. In the case of Ref. \cite{Maniscalco_2008}, the probability $P_{ee}=0$ so that dark phase is momentary, determined by (the analogue of) the probabilities of the two entangled components, whereas in the case of Ref. \cite{ficek2006dark}, the dark phase persists for finite time during which $|P_{ss}-P_{aa}|<2\sqrt{P_{ee}P_{gg}}$. A final difference between our results is that the subradiant state, and hence GME, persists asymptotically because of the DFS property, whereas in Ref. \cite{ficek2006dark} the antisymmetric state and consequent entanglement eventually die out. 
\color{black}

In the two-qubit case, the initial joint-state can be written as 
\begin{equation}
    \ket{\Psi(0)}=(\eta_+\ket{\psi_+}+\eta_-\ket{\psi_-})\ket{\bf 0}_R
\end{equation}
where $\ket{\psi_{\pm}}=1/\sqrt{2}(\ket{ge}\pm\ket{eg})$.
At time $t$ the state can be written as 
\begin{equation}
    \ket{\Psi(t)}=(\alpha(t)\eta_+\ket{\psi_+}+\eta_-\ket{\psi_-})\ket{\bf 0}_R + \beta(t)\eta_+\ket{gg}\ket{\bf 1_k}_R,
    \label{eq:case2}
\end{equation}
where $|\alpha(t)|^2 + |\beta(t)|^2=1$. At time $t^*$ such that $\alpha(t)\eta_+=\eta_-$, the $\ket{eg}$ terms in the expanded form of Eq.~(\ref{eq:case2}) cancel out, making the two system qubits separable, before subsequent evolution makes them entangled. At $t=\infty$ the state becomes
\begin{equation}
    \ket{\Psi(\infty)} = \eta_+\ket{gg}\ket{\bf 1_k}_R+\eta_-\ket{\psi_-}\ket{\bf 0}_R,
\end{equation}
which is found by direct computation to possess concurrence for various initial states.

Fig.~\ref{fig:fourplots}(c) represents the case of an initially separable state [$p=1$ in Eq.~(\ref{eq:coeffinit})] in the good cavity limit ($R=10$). As in previous instances, the solid (red), dashed (blue), and dash-dotted (magenta) curves correspond to the scenarios of maximum asymptotic entanglement, uniform coupling ($r_1=r_2=r_3=\frac{1}{\sqrt{3}}$) and $(r_1=1, r_2=r_3=0)$. In the first two cases above, as expected, $\mathcal{N}_{\rm CR}(t)$ starts from zero, acquires a finite value and reaches the steady state/stationary value through oscillations. These can be attributed to good cavity/strong coupling with the environment leading to long memory time in terms of  the reservoir correlation function $f(\tau)$, which decays slowly over time. This feature in turn leads to non-Markovian behavior where the information can flow back and forth between the system, as discussed in Section \ref{sec:revival}. In the case $r_1=1,r_2=r_3=0$, no multipartite entanglement can be created, as only one atom is interacting with the bath.

Fig.~\ref{fig:fourplots}(d) represents the case of initially $W$ state [$p=0$ in Eq.~(\ref{eq:coeffinit})] in the good cavity limit ($R=10$). As in previous instances, the solid (red), dashed (blue), and dash-dotted (magenta) curves correspond to the scenarios of maximum asymptotic entanglement, uniform coupling ($r_1=r_2=r_3=\frac{1}{\sqrt{3}}$) and $(r_1=1, r_2=r_3=0)$. 
In all cases, the tripartite negativity $\mathcal{N}_{\rm CR}(t)$ shows an oscillatory decay due to the finite corelation time of the reservoir which generates memory effect  and for some values of $r_1,r_2$ they reach a stationary value asymptotically because of the existence of non-zero overlap of the initial state with the decoherence free states (tuned by the values of $r_1,r_2$).

\section{DFS in the $n$-partite case}
\label{sec:dfsnqubit}
It is straightforward to extend the 3-particle case to that of $n$ particles. 
Explicitly, the free and interaction Hamiltonians are given by:
\begin{subequations}
\begin{eqnarray}
    H_0 &= \omega_0 \sum_{i=1}^n \sigma^{(i)}_+ \sigma^{(i)}_- + \sum_k \omega_k b_k^\dagger b_k,
    \label{eq:free_n} \\
    H_{\text{int}} &= \left( \sum_{j=1}^n \alpha_j \sigma^{(j)}_+  \right) \sum_k g_k b_k + \text{h.c.}
    \label{eq:Hint_n}
\end{eqnarray}
\label{eq:Hamiltonian_n}
\end{subequations}
In the following we shall use for convenience the notation
$$
    \ket{[j]}\equiv \ket{g}^{\otimes (j-i)}\ket{e}\ket{g}^{\otimes (n-j)}
$$

For the $n$-qubit system governed by the Hamiltonian in Eq.~(\ref{eq:Hamiltonian_n}), the subradiant states span a decoherence-free subspace (DFS) of dimension $(n-1)$. 
The argument proceeds as follows. Generalizing the three-qubit case, we construct $\binom{n}{2}$ subradiant (decoherence-free) states, orthogonal to the superradiant state  
\begin{equation}
    \ket{\psi_+^{[n]}} \equiv \sum_{k=1}^n r_k \ket{[k]},
\end{equation}
namely
\begin{equation}
\ket{\psi_{-,jk}} 
    \equiv \frac{1}{\sqrt{r_j^2 + r_k^2}}
    \left(r_k \ket{[j]} - r_j \ket{[k]}\right),
    \qquad 1 \le j < k \le n.
\label{eq:n_vectors}
\end{equation}
These states are neither mutually orthogonal nor linearly independent. 
Denote
\begin{equation}
    W \equiv \mathrm{span}\{\ket{\psi_{-,jk}}\}.
\end{equation}
Since $\ket{\psi_+^{[n]}}$ is orthogonal to every vector in $W$, the subspace $W$ satisfies
$\dim(W) \le n-1$. To determine its dimension, we identify a linearly independent subset of $W$. 
A convenient choice is the set of $(n-1)$ (non-normalized) vectors
\begin{equation}
    \ket{v_j} \equiv r_1 \ket{[j]} - r_j \ket{[1]}, 
    \qquad j=2,3,\ldots,n,
\end{equation}
each of which is proportional to $\ket{\psi_{-,1j}}$ and therefore lies in $W$. While the vectors $|v_j\rangle$ are not normalized, they are manifestly linearly independent, in that each $|v_j\rangle$ contains a unique nonzero component along the single-excitation basis vector $|[j]\rangle$.
Hence no nontrivial linear combination can vanish, and the set $\{|v_j\rangle\}_{j=1}^{n-1}$ spans an $(n-1)$-dimensional subspace. Therefore, $\dim(\mathrm{DFS}) = n-1.$ \color{black}

\section{Entanglement Revival in The Markovian Regime \label{sec:revival}}
An important observation is the revival of the entanglement observed in the context of . It is worth noting that the revival of $\mathcal{N}_{\rm CR}^*$ in Fig.~\ref{fig:fourplots}(b) or of concurrence in Fig.~4(b) in~\cite{Maniscalco_2008} occurs in the Markovian regime, and thus is not a manifestation of non-Markovian recoherence. To show this in the present situation, we restrict to the 3-qubit subspace of interest, which is span[$\{\ket{\psi^+},\ket{\psi^1_-},\ket{\psi^2_-},\ket{ggg}\}$]. The noise here naturally suggests a decomposition of the Hilbert space as:
\begin{equation}
    \mathcal{H}_{\rm total}=\mathcal{H}_{+0}\oplus\mathcal{H}_{12}.
    \label{eq:tensorsum}
\end{equation}
where $\mathcal{H}_{12} \equiv \text{span}[\ket{\psi^1_-},\ket{\psi^2_-}]$ and $\mathcal{H}_{+0} = \text{span}[\ket{\psi^+},\ket{ggg}]$. The subradiant states do not evolve with time, while the noise operator $\mathcal{E}_{\rm total}$ acts in the subspace $\mathcal{H}_{+0}$, i.e., $\mathcal{E}_{\rm total}=\mathcal{E}_{+0} \oplus \mathbb{1}_{12}$, where \(\mathbb{1}\) is the identity operation.

The noise $\mathcal{E}_{\rm total}$ can be characterized as a (collective) amplitude damping map given by the operator sum representation: 
\begin{equation}
    \mathcal{E}_{\rm total}[\rho]=E_0(t)\rho E^\dagger_0(t)+E_1(t)\rho E^\dagger_1(t),
\end{equation}
where $E_0(t)$ and $E_1(t)$ are 
\begin{align}
E_0(t) &= \begin{pmatrix}
\Phi(t) & 0 \\
0 & 1
\end{pmatrix} \oplus \mathbb{1}_{12}, 
\nonumber \\
E_1(t) &= \begin{pmatrix}
0 & 0 \\
\sqrt{1-\Phi^2(t)} & 0
\end{pmatrix} \oplus \mathbb{0}_{12}.
\label{eq:ampdamp}
\end{align}
The corresponding master equation for the reduced density matrix $\rho$ of the system is
\begin{equation}
\label{lindbladad}
\dot{\rho} = \Gamma (t) \left( 2 L_- \rho L_+ - \{L_+ L_-, \rho\} \right),
\end{equation}
where the (collective) Lindblad operator $L_- \equiv |ggg\rangle \langle \psi^+| \oplus \mathbb{0}_{12}$, $L_+ = (L_-)^{\dagger}$ and the decay rate
   $\Gamma (t) = -2\frac{\dot{\Phi}(t)}{\Phi(t)}$. 

A quantum dynamical map 
$\mathcal{E}(t)$ is said to be CP-divisible if it can be expressed as a sequence of completely positive (CP) maps, such that the intermediate maps remain CP for any time interval.
Correspondingly, all decay rates in the associated master equation are positive \cite{Rivas_2010,hall2010}. In  the bad cavity limit where $\gamma \gg \lambda \alpha_T^2$, we find $\Gamma(t)=\frac{\lambda\alpha_T^2}{\gamma}$, a positive constant indicative of a Markovian semigroup. The entanglement revival observed in Fig.~\ref{fig:fourplots}(b) (and analogously, concurrence revival in Fig. 4(b) of~\cite{Maniscalco_2008}) is thus not a manifestation of non-Markovian recurrence, but a transition between bipartite and tripartite entanglement within the Markovian regime. By contrast, the entanglement revivals in Fig. \ref{fig:fourplots}(d) reflects CP-indivisible, and hence non-Markovian, dynamics. 

These considerations generalize naturally to the $n$-qubit case. In this case, the subspace of interest is spanned by the superradiant state $\ket{\psi^{[n]}_+}$, the subradiant states  $\ket{v_2},\ket{v_3}, \cdots, \ket{v_{n}}$ and $\ket{g}^{\otimes n}$. The $n$-qubit noise determined by Hamiltonian Eqs.~(\ref{eq:free_n}) and (\ref{eq:Hint_n}) leads to a partition of the Hilbert space as
$
    \mathcal{H}_{\rm total}=\mathcal{H}_{+0}\oplus\mathcal{H}_{[n-1]}.
$
Where $\mathcal{H}_{[n-1]} \equiv \text{span}[\ket{v_2},\ket{v_3}, \cdots, \ket{v_n}]$ and $\mathcal{H}_{+0} = \text{span}[\ket{\psi^{[n]}_+},\ket{g}^{\otimes n}]$. As before, the subradiant states do not evolve with time, while the noise operator $\mathcal{E}_{\rm total}$ has the structure $\mathcal{E}_{\rm total}=\mathcal{E}_{+0} \oplus \mathbb{1}_{[n-1]}$, where $\mathcal{E}_{+0}$ is the noise given by Eq.~(\ref{eq:ampdamp}), with the same functional form of $\Phi(t)$.

Interestingly, the Markovian revival demonstrated in the 3-qubit case can be shown to generalize to the $n$-qubit case. Specifically, we find that for an $n$-qubit system governed by the Hamiltonian $H_0+H_{\rm int}$ in Eq.~(\ref{eq:Hint_n}), there exist initial states that show Markovian revival. As before, it will be convenient to assume that the initial state has a one-dimensional support in the DFS.

To show this, consider a system initially prepared in the state 
\begin{equation}
    \ket{\psi(0)}=\eta^{[n]}_+\ket{\psi^{[n]}_+}+\eta^{[n]}_-\stretchket{\psi_-^{[n]}},
\end{equation} 
where $\stretchket{\psi_-^{[n]}}$ is a subradiant state 
\begin{equation}
\label{n-subrad}
     \stretchket{\psi_-^{[n]}}\equiv\dfrac{1}{\sqrt{\bigg(\sum_{i=1}^{n-1}r_i^2\bigg)}}\big(\sum_{k=1}^{n-1} r_kr_n\ket{[k]}-\bigg(\sum_{i=1}^{n-1}r_i^2\bigg)\ket{[n]}\big),
\end{equation}
constructed to produce the following behavior. 

The time evolution under the system-bath Hamiltonian Eqs. (\ref{eq:free_n}), (\ref{eq:Hint_n}) leads to the entangled state
\begin{align}
    \ket{\psi(t)}=\Big(\alpha(t)\eta^{[n]}_+\ket{\psi^{[n]}_+}+\eta^{[n]}_-\stretchket{\psi_-^{[n]}}\Big)\ket{\bf0}_R+ \nonumber\\ \beta(t)\eta^{[n]}_+\ket{g}^{\otimes n}\ket{\bf1_k}_R,
\label{ndimevo}
\end{align}
analogous to the 3-qubit case Eq.~(\ref{eq:case2}).
There will be a time $t^*$ such that, $\alpha(t^*)\eta^{[n]}_+r_n=\eta^{[n]}_-\sqrt{\sum_{i=1}^{n-1} r_i}$,  whereby the parenthesized expression in the first term in the r.h.s of Eq.~(\ref{ndimevo}) simplifies into the biseparable form
\begin{equation}
    (\alpha(t)\eta^{[n]}_+\ket{\psi^{[n]}_+}+\eta^{[n]}_-\ket{\psi^{[n]}_-})= \ket{\psi_{\ast[n-1]]}}_{\not{n}}\ket{g}_n,
\end{equation}
where $\ket{\psi_{*[n-1]}}_{\not{n}}$ is a  $(n-1)$-party entangled state of all $n$ qubits barring the $n^{th}$ qubit which factors out at time $t^*$. For $t>t^*$, the first term in Eq.~(\ref{ndimevo}) regains its genuine $n$-party entanglement. In turn, this leads to a $n$-qubit GME for reasons analogous to that mentioned in the three-qubit case above. A general proof for $n$-qubit GME in this case is relegated to Theorem \ref{thm:nGME} in Appendix \ref{sec:npartite}, where we show that any mixture of a multiqubit $W$-class state with the ground state is entangled (Theorem \ref{thm:nME}) and in fact genuinely so (Theorem \ref{thm:nGME}) if and only if the weight of the $W$-class component in the mixture is non-zero. This ensures that the asymptotic states encountered in our dynamics are genuinely multipartite entangled.  Given a $n$-qubit pure-state GME measure $\mathfrak{G}^{(n)}$, its mixed state version by convex roof extension applied to a mixture of an $n$-qubit W-class state and the ground state can be obtained by averaging it in this same decomposition, which is shown to be optimal by virtue of Theorem \ref{thm:nGME}. During the action of the considered collective dissipative noise in the Markovian limit will produce a pattern of fall and revival analogous to that witnessed in the solid line plot in Fig.~\ref{fig:fourplots}(b).

It is worth noting that the choice of the $|\psi_-^{[n]}\rangle$ in Eq.~(\ref{n-subrad}) hinges on the minus sign of the $\ket{[n]}$ term, which is crucial to produce the momentary breakdown of GME at $t^*$. A ``wrong'' subradiant state may lack GME, as for example $\ket{\eta_-^1}$ in Eq. (\ref{eq:psi12}). However, having multiple subradiant states would require a careful analysis to keep track of the different coherences, which can potentially lead to thwarting GME death (because a cancellation between two terms is compensated by another term in the superposition) or produce multiple GME death and revival events (through cancellations at different superposition terms).\color{black}

The analysis above assumes identical qubits, collective system-bath
coupling, and a Lorentzian reservoir at zero temperature. It is useful to briefly comment on the robustness of the predicted
Markovian revival when these idealized conditions are relaxed. First, a finite bath temperature introduces thermal excitation processes in addition to spontaneous emission. While this modifies
the detailed population dynamics, the symmetry responsible for the DFS is not fundamentally affected as long as
the system-bath coupling remains collective. The revival is therefore
expected to persist qualitatively, although with reduced contrast due
to thermal population of states outside the DFS. Second, replacing the Lorentzian spectral density with an Ohmic or
related form changes the reservoir correlation function and hence
the effective decay rates. Since the revival mechanism originates
from the interplay between the radiatively coupled component of the
state and the subradiant modes within the single-excitation sector, rather than from specific features of the spectral density, the qualitative behavior should remain, although the revival time scale
and sharpness may change. Finally, weak inhomogeneity in the qubit transition frequencies breaks the exact symmetry underlying the DFS. In this case the subradiant states acquire small decay rates, so that the DFS becomes only approximately protected. The revival is therefore expected to become progressively damped as the frequency mismatch increases, eventually disappearing when the detuning becomes
comparable to the collective decay rate. A quantitative analysis of these effects lies beyond the scope of the present work but would be an interesting direction for future study.
\color{black}
\section{Discussion and Conclusions}
In this work, we investigated the dynamics of a three-qubit system collectively interacting with a zero-temperature bosonic environment characterized by a Lorentzian spectral density. Our analysis reveals several important features concerning decoherence-free subspaces, genuine multipartite entanglement evolution, and the transition between Markovian and non-Markovian regimes. First, we identified the conditions under which decoherence-free subspaces emerge, demonstrating that certain collective states remain protected from environmental dissipation because of the symmetry in system-bath interactions.  The DFS structure identified in the present
model also has concrete implications for several quantum information protocols. First, DFS-protected states can serve as robust carriers of multipartite entanglement in entanglement-based communication tasks, where maintaining correlations during transmission is essential. Second, the persistence of subradiant components suggests a natural
application to quantum memory, since information encoded in the DFS
remains immune to collective dissipation. In addition, the Markovian revival of genuine tripartite entanglement observed here may be useful in distributed quantum-network settings such as quantum repeaters, where temporary degradation of entanglement during transmission or
storage could be mitigated by intrinsic revival mechanisms without requiring active error correction. More broadly, these features highlight how collective dissipation and DFS structure can be
exploited as passive resources for stabilizing multipartite quantum
correlations in realistic open-system architectures. \color{black}These subspaces play a crucial role in preserving quantum coherence, suggesting potential applications in fault-tolerant quantum information processing. 

Next, we examined the dynamics of the genuine multipartite entanglement using the convex roof construction based on negativity, which is a pure-state genuine multipartite entanglement measure derived from the partial transpose of the density matrix. Our results highlight distinct behaviors in the Markovian and non-Markovian regimes. In the non-Markovian regime, memory effects in the bath lead to entanglement revival, where quantum correlations reappear after initial decay. Surprisingly, even in the Markovian regime, we observe transitions between genuine tripartite and bipartite entanglement, with instances of partial revival due to collective qubit-bath coupling. This behavior underscores the complex interplay between system symmetries and environmental noise, particularly in structured reservoirs. Further, we showed that our results on decoherence free subspaces and genuine multipartite entanglement dynamics for the 3-qubit system generalize straightforwardly to the $n$-qubit case.

Our findings contribute to a deeper understanding of open quantum systems and the conditions under which entanglement can be stabilized or recovered. The persistence of coherence in DFS and the observed transitions between different entanglement regimes offer insights for quantum error mitigation and the design of robust quantum memory architectures. Future work could explore engineered environments or time-dependent couplings to further enhance entanglement preservation in practical quantum devices.  Here hybrid systems that combine Zeno effect and DFS to protect entanglement would be worth investigating. Another future direction would be to revisit our results with numerical optimization to explicitly compute the convex roof extension of negativity measure in place of the use of its upper bound.

\acknowledgements
SG acknowledges the financial assistance from the Udupi Sri Admar Mutt Education Foundation. RS acknowledges partial financial support of the Indian Science \& Engineering Research Board (SERB) grant
CRG/2022/008345.

\bibliography{threequbit}

\appendix

\section{Genuine $n$-partite entanglement of asymptotic post-revival state \label{sec:npartite}}
In the $n$-qubit case, in place of Eq. (\ref{eq:canonical0}), the asymptotic state is of the form:
\begin{equation}
    \rho_W = p \ketbra{\psi_W}{}{\psi_W} + (1-p)\ket{g}^{\otimes n}\bra{{g}}^{\otimes n},
    \label{eq:canonicaln}
\end{equation}
with $p \in [0, 1]$, where $|\psi_W\rangle \equiv \sum_{i=1}^n \alpha_i |[i]\rangle$ ($\forall_j \alpha_j\ne0$) is a $W$-class state.  

For completeness, we first we demonstrate using the PPT criterion that the state Eq. (\ref{eq:canonicaln}) is entangled for $p>0$, without ruling out biseparability. As such, this proof of entanglement (rather than GME) works even if only two of the $\alpha_j$'s are non-vanishing.
\begin{proposition}
The state Eq. (\ref{eq:canonicaln}) is entangled if and only if $p > 0$.
\label{thm:nME}
\end{proposition}
\begin{proof}
To prove entanglement, we employ the Peres-Horodecki (PPT) criterion, by which a state is entangled if its partial transpose ($\rho^{T_1}$, etc.)  possesses a negative eigenvalue (but the converse is not true). First consider the ``if'' direction. Performing the partial transpose on the first qubit, the mapping $\ket{i_1, i_{rest}}\bra{j_1, j_{rest}} \to |j_1, i_{rest}\rangle\langle i_1, j_{rest}|$ transforms the coherence terms of the $W$-state. Specifically, the term $p\alpha_1 \alpha_2^* |eg\dots g\rangle\langle ge\dots g|$ becomes $p\alpha_1 \alpha_2^* |gg\dots g\rangle\langle ee\dots g|$.
Consider the $2 \times 2$ principal submatrix $M$ of $\rho^{T_1}$ indexed by the basis vectors $\{|gg\dots g\rangle, |eeg\dots g\rangle\}$. The entries are:
$ M = \begin{pmatrix} 
\langle gg\dots g| \rho^{T_1} |gg\dots g\rangle & \langle gg\dots g| \rho^{T_1} |eeg\dots g\rangle \\
\langle eeg\dots g| \rho^{T_1} |gg\dots g\rangle & \langle eeg\dots g| \rho^{T_1} |eeg\dots g\rangle 
\end{pmatrix} = \begin{pmatrix} 
1-p + p|\alpha_1|^2 & p\alpha_1\alpha_2^* \\
p\alpha_1^*\alpha_2 & 0 
\end{pmatrix}. $
The $(2,2)$ entry is zero because the state $|eeg\dots g\rangle$ has a zero population in the original mixture $\rho$.
The characteristic equation $\det(M - \mu I) = 0$ yields:
$\mu^2 - (1-p + p|\alpha_1|^2)\mu - p^2|\alpha_1\alpha_2|^2 = 0. $
The smallest eigenvalue $\mu_{\rm min}$ of this submatrix is:
$ \mu_{\rm min} = \frac{(1-p + p|\alpha_1|^2) - \sqrt{(1-p + p|\alpha_1|^2)^2 + 4p^2|\alpha_1\alpha_2|^2}}{2} $
Since $4p^2|\alpha_1\alpha_2|^2 > 0$ for all $p > 0$, it follows that $\mu_{\rm min} < 0$.

By the \textit{Cauchy Interlacing Theorem}~\cite{Horn1985}, the eigenvalues $\lambda_k$ of the full $2^n \times 2^n$ Hermitian matrix $\rho^{T_1}$ interlace with the eigenvalues $\mu_k$ of its $2 \times 2$ principal submatrix $M$. Specifically:
$$ \lambda_{\rm min}([\rho_{[n]}(\infty)]^{T_1}) \le \mu_{\rm min}(M).$$
Because $\mu_{\rm min} < 0$, the full matrix $\rho^{T_1}$ must have at least one negative eigenvalue $\lambda < 0$. Thus, by the PPT criterion, the state $\rho_{[n]}(\infty)$ is entangled for all $p \in (0, 1]$.

Consider the ``only if'' direction. If the state in the family Eq. (\ref{eq:canonicaln}) yields a negative eigenvalue, it cannot be $\ket{ggg\cdots g}$, which is a product state. It follows that it must be family member with $p>0$.
\end{proof}

The following theorem strengthens the above result to establish the GME of the state Eq. (\ref{eq:canonicaln}). It generalizes Theorem \ref{thm:3GME} to the $n$-partite case.
\begin{theorem}
The state Eq. (\ref{eq:canonicaln}) has GME if and only if $p > 0$.
\label{thm:nGME}
\end{theorem}
\begin{proof}
The ``if'' direction ($p>0$): The restriction of $\rho_W$ to the single–excitation subspace
$
\mathcal{H}_1=\mathrm{span}\{|[1]\rangle,\cdots,|[n]\rangle\}.
$
yields the pure but subnormalized state
$
\rho_1 = p\ketbra{\psi_W}{}{\psi_W}.
$
Hence $\rho_1$ is a rank $1$ matrix and therefore
$
\det(\rho_1)=0$.

We consider any bi-separable state separable across a bipartition $S|\bar S$, where $|S|=m<n$ and $|{\bar S}|=n-m$.
Because the global state has at most one excitation, a product state across this cut cannot simultaneously populate both sides (otherwise double excitations would occur).
Thus its single–excitation density matrix has support entirely within one side, implying
$$
(\rho_{S|\bar S})_{ij}=0
\quad\text{whenever } i\in S,\ j\in\bar S .
$$
Therefore the single–excitation matrix of any biseparable state,
being a convex mixture of such contributions, has the form
$
\rho_{\mathrm{bisep},1}=\sum_k q_k M_k,
$
where each $M_k$ contains zeros in at least one set of off–diagonal positions.
In order to reproduce the target matrix $\rho_1$, whose entries
$
(\rho_1)_{ij}=p a_i a_j^* \neq 0 \quad (i\neq j)
$
are all nonzero, contributions from different bipartitions must be mixed.
Such a mixture necessarily yields
$
\mathrm{rank}(\rho_{\mathrm{bisep},1}) \ge 2,
$
and hence
$
\det(\rho_{\mathrm{bisep},1})>0.
$
This contradicts $\det(\rho_1)=0$. Therefore $\rho$ cannot be written as a mixture
of biseparable states and is genuinely multipartite entangled. 

The ``only if'' direction ($p=0$): In this case, $\rho_W=\ketbra{ggg\cdots g}{}{ggg\cdots g}$, which is a product state and manifestly lacks GME.
\hfill
\end{proof}
Theorem \ref{thm:nGME} implies that the decomposition Eq. (\ref{eq:canonicaln}) is optimal to quantify GME using any pure-state measure of GME via a convex roof construction. Therefore, given ${\mathfrak G}$, such a pure-state measure
\begin{equation}
\mathfrak{G}(\rho_{[n]}(\infty)) = p\mathfrak{G}(\ketbra{\psi_W}{}{\psi_W}).
\end{equation}
This generalizes the corresponding result for the 3-qubit case.

\end{document}